\documentclass[structabstract]{aa} 
\usepackage{epsfig}
\usepackage{graphicx}
\usepackage{longtable,lscape}
\usepackage[usenames]{color}
\usepackage{ulem}

\newcommand{\nc}{\newcommand}
\nc{\Teff}{$T_{\rm eff}$\,}
\nc{\logg}{log\,$g$\,}
\nc{\kms}{\,${\rm km.s}^{-1}$\,}
\nc{\Msun}{$M_{\odot}\, $}
\nc{\Mcz}{$M_{CZ}\ $}
\nc{\vsini}{$v \sin i$}
\nc{\vrad}{$v_{\rm rad}$}
\nc{\Massloss}{\rm $M_{\odot}$.yr$^{-1}$\,}
\nc{\Lsun}{$L_{\odot}\ $}

\usepackage{txfonts}
\begin{document}

\title{Search for surface magnetic fields in Mira stars.  
\thanks{Based on observations obtained at the T\'elescope Bernard Lyot (TBL, USR5026) at the Observatoire du Pic du Midi, 
operated by the Observatoire Midi-Pyr\'en\'ees, Universit\'e de Toulouse (Paul Sabatier), 
Centre National de la Recherche Scientifique of France. }}
 \subtitle{First detection in  $\chi$ Cyg. }

\author{                        A. L\`ebre                                                                \inst{1}
\and 					M. Auri\`ere										 \inst{2,3}
\and                              N. Fabas                                                               \inst{4,5}
\and                                D. Gillet                                                           \inst{6} 
\and                                F. Herpin                                                           \inst{7} 
\and                                R. Konstantinova-Antova                                     \inst{8,3} 
\and 					P. Petit										 \inst{2,3}
}

\offprints{A. L\`ebre\\
 Agnes.Lebre@univ-montp2.fr}

\institute{                   LUPM - UMR 5299 - CNRS and Universit\'e Montpellier II - Place E. Bataillon, 34\,090 Montpellier, France 
\and				CNRS, Institut de Recherche en Astrophysique et en Plan\'etologie, 14 Avenue E. Belin, 31\,400 Toulouse, France 	
\and 			Universit\'e de Toulouse, UPS-OMP-IRAP, Toulouse, France 	  
\and			  Instituto de Astrof{\'i}sica de Canarias, 38\,205, La Laguna, Tenerife, Spain
\and 			 Departamento de Astrof{\'i}sica, Facultad de F{\'i}sica, Universidad de la Laguna, Tenerife, Spain
\and			  Observatoire de Haute-Provence - OSU Pyth\'eas - CNRS and Aix-Marseille Universit\'e,  04\,870 Saint Michel l'Observatoire,  France
\and                       LAB - UMR 5804 - OSU  Bordeaux - CNRS and Universit\'e de Bordeaux – 2, rue de l'Observatoire, 33\,271 Floirac, France
\and                    Institute of Astronomy, Bulgarian Academy of Sciences, Sofia - Bulgaria 
}

\abstract
 {So far, surface magnetic fields have never been detected on Mira stars. Only recently have the
spectropolarimetric capabilities of measuring it via the Zeeman effect become
available to us. Then, to complete the knowledge of the magnetic field and of its influence during the transition from asymptotic giant branch to 
planetary nebulae stages, we have undertaken a search for magnetic fields on the surface of Mira stars. }  
 {Our main goal is to constrain - at this stage of stellar evolution - the surface magnetic field (presence 
and strength) and to define the magnetic field strength dependence along the radial distance to the star, above the photosphere and across the circumstellar envelope of cool and evolved stars.}
 {We used spectropolarimetric observations (Stokes V spectra probing circular polarization), collected with the Narval instrument at TBL, in order to detect - with the least squares deconvolution (LSD) method - a Zeeman signature in the visible part of the spectrum. }
 {We  present the first spectropolarimetric observations of the S-type Mira star $\chi$ Cyg, performed around its maximum light. We detected a polarimetric signal in the Stokes V  spectra and established its Zeeman origin. We claim that it is  likely to be related to a weak magnetic field present at the photospheric level and in the lower part of the stellar atmosphere.  
We estimated the strength of its longitudinal component to about 2-3 gauss. This result favors a  $1/r$  law for the variation of the magnetic field strength  across the 
circumstellar envelope of $\chi$ Cyg. This is the first detection of a weak magnetic field on the stellar surface of a Mira star, and we discuss its origin in the framework of shock waves periodically propagating throughout the atmosphere of these radially pulsating stars. 
}
 {At the date of our observations of $\chi$ Cyg, the shock wave reaches its maximum intensity, and it is likely that the shock amplifies a weak stellar magnetic field during its passage through the atmosphere. 
Without such an amplification by the shock, the magnetic field strength would have been too low to be detected. 
For the first time, we also report strong Stokes Q and U signatures (linear polarization) centered on the zero velocity (i.e., on the shock front position). They seem to indicate that the radial direction would be favored by the shock during its propagation throughout the atmosphere.
}


\maketitle
\titlerunning{Search for Magnetic Fields in Mira Stars. First Detection in $\chi$ Cyg}
\authorrunning{A. L\`ebre, M. Auri\`ere, N. Fabas et al.}
\date{Received 11 October 2013,  accepted 21 November 2013}
%

\section{Introduction}

For cool and evolved stars, e.g., objects along the asymptotic giant branch (hereafter AGB),  the stellar magnetism can be one of the ingredients participating in the mass loss process. 
It can help shape the morphology of the circumstellar envelope (hereafter CSE) surrounding the object at this peculiar phase of stellar evolution, and it can also rule the CSE's evolution in the subsequent 
stages  (Blackman et al., 2001). 
While in the past decade magnetic fields have been detected and measured throughout the circumstellar environments of cool and evolved stars 
(see Vlemmings, 2011) and also on the surface of late type giants  (see Konstantinov-Antova et al., 2013), surface magnetic fields have so far never been reported on Mira stars.\\

Miras are cool and evolved pulsating stars that belong to the AGB, the key evolutionary stage of an 
intermediate mass star before its transition toward the planetary nebulae (hereafter PN) stage. These variable stars are known to be the 
main recycling agents of the interstellar medium since they undergo a prodigious mass loss  supposedly mainly driven by radiation pressure on dust 
(see for example Willson, 2000). However a combination of several factors is expected to play an important role in the mass loss 
process, including pulsation (and its associated complex atmospheric dynamics produced by shock waves), condensation and opacity 
of dust grains in the upper atmosphere, which are still debated (see Hoefner 2011). \\

Mira stars undergo radial pulsations that lead to a rather periodic cycle of visual light variations, with a  period of about one year 
(see Willson \& Marengo, 2012).
Concerning the atmospheres of Mira stars, early observational works and, more recently, high-resolution spectroscopic studies (see, for example, Gillet et al., 1983) and 
envelope tomography  (Alvarez et al., 2000, 2001) have reported strong emissions of the hydrogen lines lasting up to 80 \% of the luminosity period. 
Those works have established that radiative and hypersonic shock waves, which are triggered by the pulsation mechanism, were periodically propagating  throughout the stellar atmosphere, generating emission lines formation process
and favoring the doubling of metallic lines. This complex atmospheric dynamics also contributes to forming and to enriching a CSE. \\

 Recently, from spectropolarimetric data (focusing on linear polarization), Fabas et al. (2011) have 
 characterized the shock wave propagation throughout the stellar atmosphere of the prototype of oxygen-rich Mira stars: Omicron Ceti. 
They report  signatures in Stokes Q \& U but also in  Stokes V parameters (tracing linear and circular polarization, respectively), 
associated to the strong Balmer hydrogen emissions known to be formed in the radiative wake of 
the shock wave (Fadeyev \& Gillet, 2004). The origin of these spectropolarimetric signatures points to global asymmetry (at least partly photospheric) 
owing to the passage of the shock's front 
throughout  photospheric giant convective cells, but this cannot exclude the presence of a weak magnetic field that still has to be detected on the surface of Miras.\\

Magnetic fields have indeed been detected and measured throughout the CSE of AGB stars. Among these,  M type AGB stars have been probed at different depths from  the polarization of the maser emission of several molecules (OH, H$_{2}$O, and SiO), located at different distances from the central star (respectively, at 1~000-10~000 au, at a few 100 au, and at 5-10 au, with one stellar radius R$_{*}$ $\sim$ 1 au). The current status of a magnetic field  throughout the CSE of an AGB star is the following 
(where B$_{//}$ is  the mean value of the strength of the magnetic field along the line of sight):\\
\noindent  B$_{//}$ $\sim$ 5-20 mG, at 1~000-10~000 au (from OH masers; Rudnitski et al., 2010), 

\noindent  B$_{//}$ $\sim$ a few 100 mG, at a few 100 au (from H$_{2}$O masers ; Leal-Ferreira et al., 2012 ), 

\noindent  B$_{//}$ $\sim$ a few gauss (mean value = 3.5 G), at 5-10 au (from SiO masers; Herpin et al., 2006), where these values represent the innermost detections  to date of a magnetic field in the environment of Mira stars.\\

 When considered together, those data point to a clear decrease of the magnetic field strength throughout the CSE of an AGB. Moreover, for a few C-rich objects, magnetic fields have been detected in the outer region of their CSE when studying the Zeeman effect in CN line emission (Herpin et al., 2009). These constraints seem to favor the strength of magnetic fields decreasing along a 1/r law throughout the regions of the CSE  (see Fig.~1 in Vlemmings, 2011). Extrapolating this law for a toroidal magnetic field configuration (as proposed by Pascoli, 1997), one can expect a magnetic field strength of a few gauss at the photosphere of Miras.\\

 Today the detection of a weak magnetic field on the surface of Mira stars seems to be reachable with the spectropolarimeter Narval installed at the 
2m {\it T\'elescope Bernard Lyot} (TBL). 
Indeed, weak magnetic fields (B$_{//}$ between 1 to 10 G) have recently  been detected with Narval on the surface of cool stars.  
Auri\`ere et al. (2009) report a weak field (i.e., below 1 G) on the surface of the K0III giant star, Pollux, and several G type and K type giants have 
been Zeeman-detected (Konstantinova-Antova et al., 2013), most of them with surface magnetic fields weaker than 10 G. Konstantinova-Antova et al. (2010) have  found several AGBs (in the spectral type range M0III - M5III) with surface fields. 
Auri\`ere  et al. (2010)  also report magnetic field detection on the surface of the red-supergiant (RSG) Betelgeuse (spectral type M2I), 
with an average strength of the order of 1 G. \\

 We  therefore undertook a search  for magnetic fields on the surface of  Mira stars through dedicated spectropolarimetric campaigns. 
In this paper we present our first spectropolarimetric observations of the S-type Mira, $\chi$ Cyg, performed with the Narval instrument. 
In Sect.\,2 we  present our main target and in Sect.\,3 we describe the observations and the data reduction process. 
The detection of a polarimetric signal in the Stokes V spectra is presented in Sect.\,4, and its physical origin is investigated  in Sect.\,5. 
Finally, a discussion is presented in Sect. \,6 and some concluding remarks given in Sect.\,7.

\section{The S-type Mira star  $\chi$ Cyg }

While 80 \% of  the observed Miras are oxygen-rich,  M spectral-type objects  (with abundance ratio 
C/O $<$1 and with TiO bands as the main atmospheric absorbers),  carbon-rich Miras of the C spectral type (C/O $>$ 1) exhibit features from C$_{2}$, CH, 
and CN molecules in their spectra. 
The peculiar intermediate S spectral type (with C/O $\sim$1) corresponds to Miras hosting TiO and ZrO and VO molecules in their environment (see Van Eck \& Jorissen, 1999). 
The S-type Mira star, $\chi$ Cyg is a variable star with a  period of pulsation of 408 days and -from maximum to minimum light-  a very high visual amplitude (more than 10 in V magnitude)  and  a spectral type varying from S6.2 to S10.4 (Samus et al., 2012).   
Its low surface temperature ($\sim$ 3600~K, around the maximum light, see for example Wallerstein, 1985) 
allows for the existence of  molecules in its very extended atmosphere and in its CSE. They  are at the origin of the strong molecular absorptions 
reported in the visible and near infrared spectrum of $\chi$ Cyg (Maehara, 1968 ; Hinkle et al., 1982 ;  Wallerstein, 1985), because they are even more present around the minimum light 
(coolest phase, at $\phi$ $\sim$ 0.5) than around the maximum light ($\phi$ $\sim$ 0.0). 
\\

Boyle et al. (1986) have performed  the first  CCD spectropolarimetric observations of $\chi$ Cyg around one of its maximum lights, 
and detailed  spectropolarimetric feature can be appreciated in intensity, fraction of linear polarization, and position angle. 
The polarization of the continuum varies from 1 \% in the blue part of the spectrum down to 0.25 \% in the red part, 
a behavior that appears frequent in Mira stars. The most striking feature is the increase in the linear polarization in the Balmer emission lines (+ $\sim$ 0.5 \%) and in the molecular bands 
(+ $\sim$ 0.5 \%), and even more significantly in the CaI line at 422.6 nm (+ $\sim$ 2 \%), while no significant change in the position angle has been reported. While all these signatures 
favor polarization arising from the stellar atmosphere (in the formation region of the spectral lines), a departure from spherical symmetry is also invoked. 
Among all the possible sources for such asymmetry,  the  complex 
atmospheric dynamics involving shock waves (Hill and Willson, 1979) has been suggested, affecting both the line formation region in the low atmosphere and the dust formation region in 
the inner part of the CSE.\\

While $\chi$ Cyg is included in their sample, Ramstedt et al. (2012) do not mention it as one of the few AGBs hosting X-ray emission.  
No rotation measurement has ever been reported on that Mira, and  no companion is known. 
From near infrared interferometric data  (with IOTA), Ragland et al. (2006) have found for this object strong evidence for a departure from circular symmetry. 
Indeed, observing $\chi$ Cyg around the phase $\phi$ $\sim$ 0.20, 
 they report a non-zero closure phase that can be interpreted in terms of an asymmetric structure close to the photosphere (or a symmetric photosphere 
with local compact features). Tenptative explanations have been proposed for this modulation of the surface brightness, such as  superficial or atmospheric processes 
including convection, magnetism, clumpy dust formation, and the interaction of a planetary or a stellar companion with the stellar wind. Later on, Lacour et al. (2009) also performed infrared interferometric imaging of $\chi$ Cyg collecting observations around and after its maximum light of 2005  
(from $\phi$ = 0.93 to  1.79). During the pulsation phase, their images of $\chi$ Cyg display important changes in the stellar diameter and in the limb darkening,  
and they also reveal stellar inhomogeneities. Model fitting of their data favors the hypothesis of a warm molecular layer located above the photosphere.
However, so far no convincing explanation has been given as to the nature or the origin of the asymmetry found in $\chi$ Cyg.\\

Ramstedt et al. (2009) have estimated mass loss rates of a sample of S-type AGB stars, including  $\chi$ Cyg. For this Mira star they derived a  distance of 110 pc 
and estimated a mass loss rate of 3.8 10$^{-7}$ \Massloss (from fitting a radiative transfer model to CO line observations).
Comparing their results with mass loss rates of M and C type AGB stars,  they argue that the same mechanism would drive the mass loss in objects of the three 
chemical types. They also find a similar dust formation efficiency in all three chemical types and no difference in their circumstellar physical properties, including circumstellar SiO abundances.\\

From the polarization of SiO maser emission, Herpin et al. (2006) have detected a magnetic field  in the inner part of the CSE of $\chi$ Cyg (i.e., at a few stellar radii from the photosphere). Considering the Elitzur model for the SiO maser theory (Elitzur, 2002), they have estimated the average magnetic field strength from the level of the circular polarization 
associated to well-identified components of the SiO maser emission profile.  From their radio observation of $\chi$ Cyg collected at 
$\phi$ $\sim$ 0.37, 
they have found  B$_{//}$ 
in the range of 0 - 8.8 G within the SiO maser emission profile, with the central and main peak (at V$_{LSR}$ = 10 \kms) associated to B$_{//}$ = 5 G. 

\section{Spectropolarimetric observations of $\chi$ Cyg and data analysis}

\subsection{Circular polarization data of 2012}

\noindent With the spectropolarimeter Narval, we  collected  a series of  circular polarization Stokes V sequences of $\chi$ Cyg around its maximum light of March 2012.  
Narval  is a twin of the ESPaDOnS instrument (Donati et al., 2006). It is a cross-dispersion echelle spectrograph with a polarimetric module 
mounted on the 2m  {\it Bernard Lyot} telescope (TBL) 
of {\it Pic du Midi} observatory, France. \\

The observations were collected from 16 to 29 March 2012, when $\chi$ Cyg was bright  (V $\sim$ 4.7). From a light curve generated with AAVSO data, a maximum light 
could be located on  5 April 2012. In fact, the maximum of the light curve is a  {\it plateau} spreading from  25 March to 12 April 2012, so that the mean phase of our observations is $\phi$ = 0.96. This specific phase represents the hottest conditions for the photosphere and for the lower part of the atmosphere, since \Teff is maximum (along the pulsation cycle), and the molecular absorption is expected to make the lowest contribution onto the visible spectrum.\\

We have obtained  a total of 174  V sequences (i.e., simultaneous Stokes I unpolarized spectra and Stokes V circularly polarized spectra) covering a large part of the visible domain 
(from 380 to 1~010 nm) with a high-resolution power (R $\sim$ 65~000). The data were reduced with  the Libre-ESpRIT software (Donati et al., 1997) 
that  performs classical operations on spectropolarimetric data 
(bias subtraction, flat-fielding, removal of bad pixels, wavelength calibration, spectrum extraction, and extraction of the polarimetric information).
Table~1 presents our  Stokes V observations, with the number of sequences collected   each night. 
UT$_{1}$ and HJD$_{1}$ refer to the precise date (respectively, in universal time  
and in heliocentric Julian date, with  HJD + 2~456~000) of the beginning of the first sequence collected on a given night, 
while UT$_{2}$ and HJD$_{2}$ refer to the precise date of the beginning of the last sequence collected (within the same given night). 
A  spectropolarimetric V sequence is composed of four successive subexposures 
and then a  complete V sequence was acquired within less than five minutes (including the CCD fast register readout time). 
 For all the V sequences, the peak signal to noise (S/N) was usually greater than 750 (around 871~nm).
Cumulating our 174 Stokes V sequences results in a total S/N of about 10~000 allowing a field detection at the sub-gauss level  using the LSD method (see Sect.\,4). 
In Table 1,  the time exposure is given in seconds and does not take  the CCD readout time into account. The last column reports the mean 
peak S/N (given per 2.6 \kms spectral bin, around 871~nm) of all the V sequences of a considered night.\\

\begin{table}
\caption{Journal of our spectropolarimetric Stokes V observations of $\chi$ Cyg, collected in March 2012. 
The date of the observations (within March 2012) is mentioned in the first column and, 
in the second column, the number of sequences (N) collected  within a given night is given. 
For a given night, UT$_{1}$ and HJD$_{1}$ refer to the precise date of the beginning of the first sequence collected during the night, 
while UT$_{2}$ and HJD$_{2}$ refer to the precise date of the beginning of the last sequence collected during this same night.
Time exposure (in seconds) does not include the CCD readout time.  The last column reports the mean 
peak S/N of all the V sequences on a given night.}
\begin{tabular}{|c|c|c|c|c|c|c|c|}
\hline

Date  & N & UT$_{1}$ & HJD$_{1}$  & UT$_{2}$ & HJD$_{2}$ &Time   & $<$S/N$>$    \\
3/12 &   & &    & &  & exp. &  \\
\hline

16 & 31 &03:23 & 2.637 &05:21 & 2.719 & 4X15 &1132\\
\hline
24 & 32 &02:32 &10.602 &05:00 &10.705 & 4X20&856\\
\hline
25 & 28 & 02:55& 11.618&04:58 &11.704 & 4X15&1296\\
\hline 
26 & 21 &03:15 &12.632 &04:40 &12.691 & 4X18&1345\\
\hline
27 & 20 &03:15 &13.632 &04:51 &13.698 &4X20 &1110\\
\hline
28 & 18 &03:09 &14.628 &04:46 &14.695 &4X18 &1075\\
\hline
29 & 24 &02:30 &15.601 &04:38 &15.690 &4X15 &1047\\
\hline
\end{tabular}
\end{table}

\subsection{Full Stokes parameters observations of 2007}

We did not obtain any linear polarization data simultaneously with our circular polarization data of 2012. However,   
we have previous Narval spectropolarimetric observations of $\chi$ Cyg  at our disposal, collected around maximum light ($\phi$ $\sim$ 0.94) on  4 September 2007 
(full Stokes parameters observations: a single Q \& a single U sequence, tracing linear polarization, plus a single V sequence). 
Table~2 presents these complementary  data, with the precise observation  date  (universal time and heliocentric Julian date), time exposure of the sequence (without CCD readout time),  and peak S/N (at 871~nm).

\begin{table}
\caption{Journal of our spectropolarimetric full Stokes observations of $\chi$ Cyg  collected on 4 September 2007 (full Stokes parameters).}
\begin{tabular}{|c|c|c|c|c|}
\hline
Stokes  & UT & HJD  & Time exp.  & Peak S/N    \\
param. & & + 2~454~300. &(s) &(871~nm)  \\
\hline
V & 20:13  &  48.347 & 4 X 120 & 2072 \\
\hline 
Q &  20:25  & 48.355  & 4 X  120&  1696\\
\hline
U &20:37  & 48.365  & 4 X 120 &  2087\\
\hline
\end{tabular}
\end{table}

\section{Detection of a Stokes V signature}\label{detec}

We  performed an LSD analysis of our data of March 2012. Using a  numeric mask, this cross-correlation technique (Donati et al., 1997) enables 
thousands of atomic lines to be averaged within a single Stokes I and  a single Stokes V profile. 
To take the  peculiarities of the stellar atmosphere of $\chi$ Cyg  into account,   
we  used  a specific numeric mask that is typical of a cool and evolved object (i.e., with \Teff = 3~500~K,  \logg = 0.5, microturbulence $\xi$ = 2 \kms 
and solar abundances). It involves about 14 000 lines (deeper than 40\% of 
the continuum) with  
atomic parameters and Land\'e factors taken from  the Vienna Atomic Line Database (VALD, Kupka et al., 1999), 
Within this numeric mask, the atomic lines are distributed  
from 380 to 1~000 nm.  They are metallic lines, formed in the lower part of the stellar atmosphere and with known Land\'e factors, as we have carefully removed the 
contributions of specific elements with 
likely emission line profiles (such as H or He) and also lines from elements tracing the upper atmosphere or the circumstellar medium (Na, Ca, K, etc.).
 \\

 When considering an individual  V sequence or when considering the average of the  V sequences collected over one night, the LSD analysis does not reveal a 
solid detection in the LSD Stokes V profile. A {\it marginal detection} is only obtained with the 
data collected on the 16 March 2012 (31 V sequences, corresponding to a total S/N of about 5~250). 
However, when  combining the 174 Stokes V sequences through the co-addition of their associated LSD profiles (weighted by their respective noise), 
the  LSD statistics report a {\it definite 
detection} (with  $\chi$$^{2}$=1.81  and false alarm probability of 5.2 10$^{-10}$). 
We recall that the different  detection flags from LSD are DD = {\it definite detection} (false alarm probability, fap $<$ 10$^{-5}$), 
MD = {\it marginal detection} (10$^{-5}$ $<$  fap $<$ 10$^{-3}$), and ND = {\it no detection} (10$^{-3}$ $<$  fap). 
Figure 1 displays the LSD profiles: 
the Stokes I and Stokes V profiles, and also the null profile (N), a diagnosis that does not contain any 
signal within the Stokes I and the Stokes V profiles. 
All the profiles are presented along the stellar rest frame velocity (from thermal microwave emission, and the star's centre of mass velocity is -7.5 \kms, Wallerstein, 1985).
A weak Stokes V signature (10$^{-5}$ amplitude level) is clearly detected on the resulting LSD Stokes V profile. 
The null profile remains flat within the line, confirming  the detection of a significant signal in Stokes V, associated to the I profile. 
In fact, the LSD Stokes I profile presents the typical  doubling of metallic lines (Schwarzschild, 1952) 
due to a shock wave imprinting complex ballistic motions in the lower part of the stellar atmosphere.
The LSD Stokes V signature appears to only be associated  to the blue component of the LSD Stokes I profile (discussed in Sect.~5.2).\\ 

\begin{figure}[h]
\includegraphics[scale=0.35,angle=270]{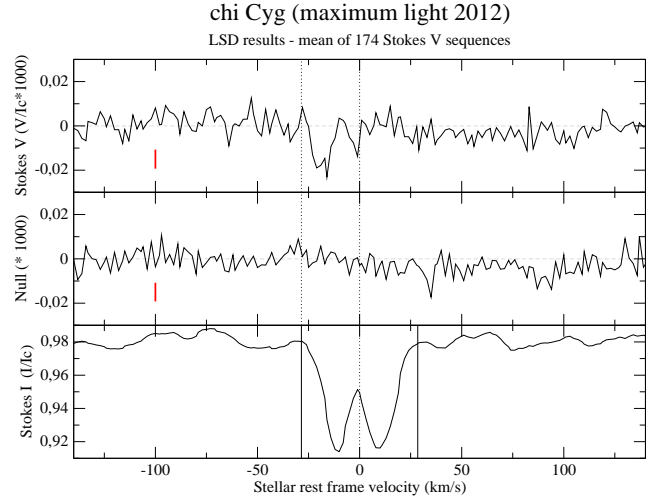}
\caption{LSD profiles of $\chi$ Cyg (in stellar rest frame velocity) from  Narval observations collected in March 2012 (average of 174 V sequences). 
{\it Bottom:} Stokes I (the unpolarized spectrum). 
{\it Middle:}  Null profile (extended by a factor of 1~000) and its associated $\pm$ 1 $\sigma$ error bar (on the left side).  
{\it Top:} Stokes V profile (extended by a factor of 1~000) and its associated $\pm$ 1 $\sigma$ error bar (on the left side).
In all panels, the vertical dashed lines help to appreciate the  location and contribution of the blue line component of the Stokes I profile.  
 The two straight lines reported around the Stokes I profile help to delimit the velocity domain used for the determining of B$_{//}$ (see Sect.~5.1).}
\end{figure}

 However weak, this Stokes V signal is statistically significant. Indeed it remains present (and is located at the same position) 
if we split our data set into two series of equal significance 
while no signature  appears in the corresponding  null-polarization profiles.  
We  also  
searched carefully for the possible contamination of the Stokes V 
channel by the  linear polarization mode and, to do so, we  used the data collected on $\chi$ Cyg at the maximum light of 2007 (see Sect.~3.2). From the LSD analyses of these data (all  individual sequences), a strong feature  is clearly detected in the Stokes U  (polarization level of about 5 10$^{-4}$) and also in the Stokes Q spectra  (polarization level of about 4 10$^{-4}$),  while their associated null profiles remain flat within the line confirming  the detection of significant signals. No  signature is found in the single Stokes V sequence (S/N $\sim$ 2\,000), while the noise level is estimated at about 0.5 10$^{-4}$. 
Figure 2 displays the LSD results 
for the September 2007 data. 
The Stokes I profile also presents the typical doubling of metallic lines (already mentioned from Fig.~1 on the 2012 data). However, its blue component 
has not yet developed fully. This is due to the slight phase lag in between our two observational sets (2007 and 2012) and also to the well known non-strictly 
periodic reproductibility of pulsation cycles within Miras.\\

\begin{figure}[h]
\includegraphics[scale=0.35,angle=270]{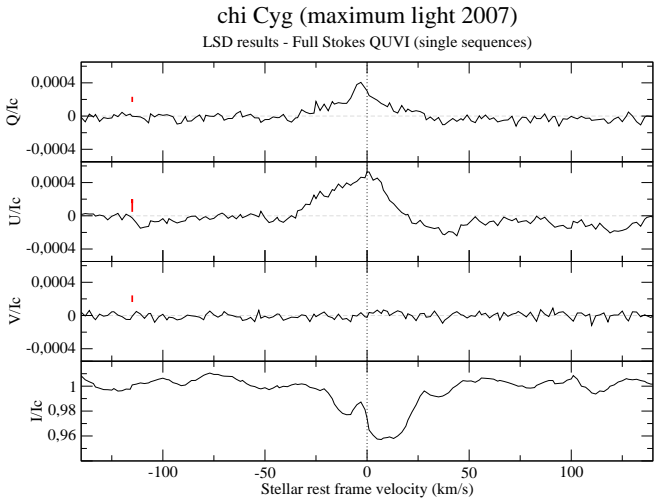}
\caption{LSD profiles
of $\chi$ Cyg (in stellar rest frame velocity) from  Narval observations collected in September 2007 (single polarimetric sequences). 
From bottom to top:  Stokes I profile (unpolarized spectrum),  Stokes V profile (tracing the circular polarization), and  Stokes U and Stokes Q profiles (tracing the linear polarization).
}
\end{figure}

While  crosstalk effects in Narval are estimated to about 1 \%, the linear polarization level detected in Stokes Q \& U spectra  ($\sim$ 4-5~10$^{-4}$) could contaminate the Stokes V spectrum  at a $\sim$ 5~10$^{-6}$ level. 
Moreover, a contamination by the linear channel would produce - within the circular channel - a V signature that presents the same configuration as the Q \& U signals:
(1) they are almost centered on the 0 \kms stellar rest frame velocity; (2) using two submasks (split into two parts - with a cut at 540~nm - the initial mask involved in our LSD analysis), they are measured to be significantly stronger in the blue domain than in the red one. 
This behavior is consistent with the observed linear polarimetry owing to scattering  in small particles (e.g., Rayleigh or coherent scattering). 
Some contamination by the linear polarization of the continuum would have the same color effect (Boyle et al.  1986), but it is not expected to be significant 
since Narval is specifically designed to trace the polarization within the spectral lines.    
Figure 3 displays the results of the LSD analysis 
performed with these two blue and red submasks on the averaged 174 Stokes V sequences collected in March 2012. Both analyses lead to  V signals  with equivalent amplitudes and the same location as with the initial and complete numeric mask (see Fig.~1). 
Indeed, when using the blue mask and although the S/N  is  poor in this part of the spectrum of a cool and evolved star, the signature is already present but very noisy.
While when using the red mask (gathering only 2~000 atomic lines), the signature is pronounced and the LSD statistics results in a {\it definite detection}, since the S/N is  higher in the red part of the spectrum.
The V signature reported from our data collected in 2012 exhibits a $\sim$ 10$^{-5}$ level amplitude (i.e., $\sim$ twice the eventual crosstalk level), and it is  clearly shifted onto the blue component  of the Stokes I 
profile (see Fig.~1). Also, it is very visible in the LSD analysis performed with the blue submask (despite the low S/N), but not stronger 
than in the one performed with the red submask. These facts rule out a significant contamination of the circular mode by the linear one.\\

\begin{figure}[h]
\includegraphics[scale=0.36,angle=270]{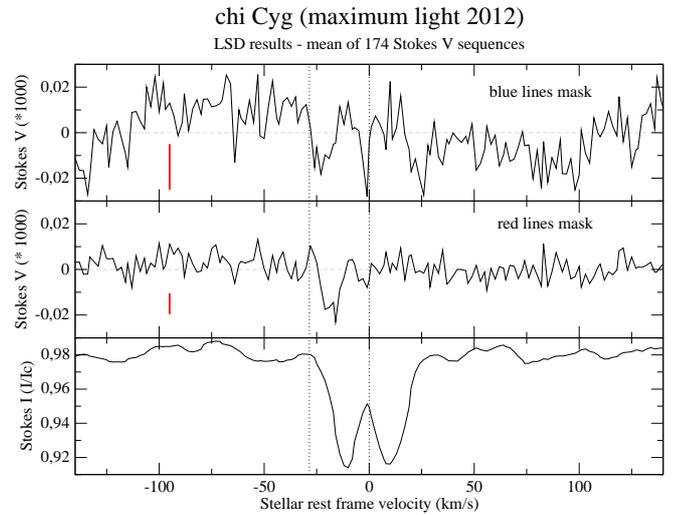}
\caption{LSD  Stokes I  and Stokes V profiles of $\chi$ Cyg (in stellar rest frame velocity) from  Narval observations collected in March 2012. 
{\it Bottom:}  LSD  Stokes I profile (same as in Fig.~ 1). 
{\it Middle:} LSD Stokes V  profile (extended by a factor of 1~000) obtained with a mask that gathers only red  lines  ($\lambda$ $>$ 540~nm).  
{\it Top:} LSD Stokes V  profile (extended by a factor of 1~000) obtained with a mask that gathers only blue  lines  ($\lambda$ $<$ 540~nm).  
}
\end{figure}

Consequently, the Stokes V signature displayed in Fig.~1 is statistically significant and related to a stellar effect (present in the region of lines formation). 
However, it is a weak signal and is only revealed  when averaging all the 174 Stokes V sequences  collected in March 2012. 

\section{Physical origin of the Stokes V signature detection  }\label{inter}

\subsection{A Zeeman (magnetic) origin}

We have established the Zeeman origin of Stokes V signature displayed in Fig.~1 by performing LSD analysis with two numeric masks issued from the initial one and by 
gathering atomic lines with respectively high and low Land\'e factors (i.e., greater or lower than the value 1.2).  Both masks contain about 7~000 atomic lines with a similar distribution over the full 
spectral range and a similar mean wavelength. 
Figure~4 displays the results of the LSD analysis performed with masks composed of atomic lines selected from their Land\'e factors. 
The LSD statisics performed with the mask that gathers only lines with low Land\'e factors (i.e., with mean value of 0.9) provides {\it no detection}. 
In contrast,  the LSD  analysis done  with the mask gathering  lines with high Land\'e factors (i.e., with mean value of 1.5)  results in a  {\it definite detection}, and a clear signature is  observable in the V spectrum.\\

\begin{figure}[h]
\includegraphics[scale=0.36,angle=270]{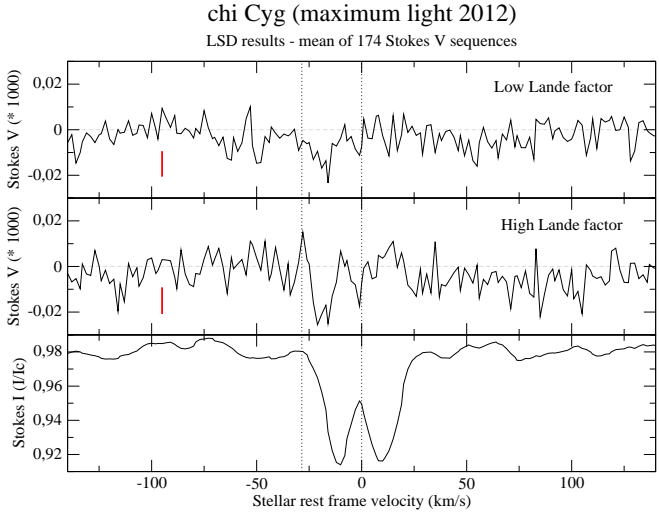}
\caption{LSD Stokes I and V  profiles of $\chi$ Cyg (in stellar rest frame velocity) from  Narval observations collected in Mars 2012 (average of 174 V sequences). {\it Bottom:} LSD 
Stokes I profile (same as Fig.~1). 
 {\it Middle:} LSD Stokes V profile obtained when using a numeric mask that only gathers atomic lines with low values of Land\'e factor (mean value = 0.9). 
{\it Top:} LSD Stokes V profile obtained when using a numeric mask that only gathers atomic lines with high values of Land\'e factor (mean value = 1.5).   }
\end{figure}

 Consequently, the Stokes V signature displayed in Fig.~1 seems to have a stellar origin sensitive to a (magnetic) Zeeman effect. 
We claim that the V signature we  found  is likely to be related to a weak magnetic field present at the photospheric level and in the lower part of the stellar atmosphere. From the LSD Stokes I and Stokes V profiles and using the first-order moment method (Rees \& Semel 1979) adapted to LSD, 
we  derived the  averaged longitudinal magnetic field associated to this signature:  B$_{//}$ = - 0.25 $\pm$ 0.40 G. 
 We note  that B$_{//}$ is the projection of the magnetic field along the line of sight and integrated over the stellar surface, so that the obtained value is lower than the field modulus. Moreover, this estimation of the longitudinal magnetic field has to be 
considered carefully. Indeed, the obtained B$_{//}$ value is 
weak (compared to its associated error) mainly because the Stokes V signature does not appear antisymmetric regarding the center of the Stokes I profile. 
Moreover this B$_{//}$ estimation is probably a lower limit (see next section) because its computation domain takes the complete width of the Stokes I profile  into account 
(i.e., blue and red components,  see the computational limits indicated in Fig.~1 with vertical straight lines).

\subsection{A link with the periodic shock wave}

As is  obvious from Fig.~1 (and Fig.~2), the LSD Stokes I profile presents the typical doubling of metallic lines due to the presence of a radiative shock 
wave  in the lower atmosphere, around the phase of the maximum light. Such a shock wave imprints ballistic motions separating atmospheric layers into two different velocity streams (Schwarzschild, 1952). The blue line component is formed by the contribution 
of  atoms located  behind the shock front and then moving outward.  The red line component is due to atoms
 located above the shock front and descending  after the perturbation of a previous shock.
Thus, the Stokes V signal is associated to 
the material that is driven outward by the shock wave. This suggests that its origin or its occurrence 
is very likely  linked to the shock: either it may be directly produced within it, or more likely, it may be due to the amplification by the shock of the weak photospheric magnetic field. \\

Estimation of the strength of the  longitudinal magnetic 
field would then have to consider  that the Stokes V profile does not present the same width as the complete Stokes I profile. 
As a result, the longitudinal magnetic field estimated above (see Sect. 5.1) could appear as a lower limit, since its computation takes the complete Stokes I profile into account. 
Scaling  the Stokes V signal and  the blue component of the I profile  obtained on $\chi$ Cyg directly to the profiles obtained from a non-pulsating K0III star (e.g., Pollux, 
hosting a classical 
Zeeman profile, see Fig.~1 in Auri\`ere et al., 2009), the V signature we report here for $\chi$ Cyg would then correspond to the detection of a surface magnetic field with 
B$_{//}$ of 2 to 3  G. This result is in good agreement with the measurement of the magnetic field (longitudinal component B$_{//}$) detected in the inner part of the CSE of $\chi$ Cyg from SiO maser 
(Herpin et al., 2006). It also clearly favors a  r$^{-1}$ variation law across the CSE, assuming a toroidal field, because its extrapolation to the photosphere of Miras leads to a surface field of few gauss. 
The r$^{-1}$ dependence law (and also the  r$^{-2}$  law, assuming a poloidal field) is  in fair agreement with the values reported throughout all  parts of the CSE of an AGB 
(see Vlemmings et al., 2011 ; Leal-Ferreira et al., 2013). However, extrapolated to the photosphere, the r$^{-2}$  law would lead to a surface field strength on the order of 
10 G, and thus it cannot be completely excluded by our observations that are limited to only one (S-type) Mira observed around its maximum light.\\

Among the data of $\chi$ Cyg collected in 2007, the LSD Stokes Q and Stokes U profiles (see Fig.~2) reveal a striking phenomenon. 
They both display a strong signature  centered 
onto the shock-front position separating the two Stokes I components, i.e., very close to the star's center-of-mass velocity. 
The location of these Q and U signatures differs from the one of the V feature appreciated in our 2012 data (see Fig~1). This leads to the linear polarization within 
the atomic lines  having a different origin than the circular polarization with the same contributing lines. An overpolarization of these metallic lines 
(with respect to their nearby continuum) would be likely to occur, owing to the contribution of a dynamic effect onto the Q and U signatures, and then the shock wave, emerging from 
the photosphere slightly before the maximum light (at $\sim$  $\phi$ = 0.8, see Wallerstein, 1985), would be the best responsible agent. 
Indeed, the peak of the Stokes Q and Stokes U signatures, centered on the zero velocity of the shock front,  
seems  to indicate that the radial direction  would be favored by the shock wave at the photospheric level, inducing a peculiar geometry, hence  the net linear polarization 
resulting after a scattering process. For $\chi$ Cyg, such an asymmetry at the photospheric level  has already  been reported  by Ragland et al. (2006) from interferometric data (see Sect.~2). 
It is also consistent with the conclusions of Fabas et al. (2011) that suggests that, in the lower part of the atmosphere of the M-type Mira Omi Cet, the interaction of the shock front 
with giant convective cells would result in the generation of a complete asymmetry. This  would prevent the linear polarization (created from the
scattering of light into the atmosphere) to be cancelled by the lack of resolution of the star. 
Therefore, the amount of linear polarization that Fabas et al. (2011) have detected in the Balmer emission lines (of Omi Cet)  is 
clearly correlated with the pulsation phase. It is worth noting that similar behavior for the Balmer emission lines and (phase variating) associated linear polarization 
have been also reported in S-type Mira stars (Fabas, 2011), including $\chi$ Cyg (Fabas, private communication).\\

Finally, the shape of the Stokes V signal reported in our 2012 data of $\chi$ Cyg also favors this link with the periodic shock wave. 
Although the high noise level compared to the amplitude of the Stokes V signatures (see Figs.~1, 3, \& 4) 
prevents a detailed investigation of the shape of this circularly polarized signal, there is marginal evidence for the presence 
of two positive lobes and two negative lobes. There is also marginal evidence that the negative lobes are more prominent than the positive ones 
(resulting in a non-zero integral of the Stokes V signature), and  this complete Stokes V asymmetry is not expected in the standard theory of the Zeeman effect. 
However, this phenomenon has been well-documented for the Sun (e.g., Viticchi\'e \& Sanchez Almeida, 2011), where 
it is interpreted as the result of combined magnetic and velocity gradients in the solar photosphere 
(Lopez Ariste, 2002). 
Stokes V asymmetry is also reported for a few cool active stars, where a high S/N  reveals this effect quite prominently (Petit et al., 2005; 
Auri\`ere et al., 2011; Petit et al., 2013). Most Zeeman signatures reported for cool supergiants (Grunhut et al., 2010) 
exhibit the same effect, which may be linked to sharp velocity/magnetic gradients related to the shock waves generated by supersonic convection 
(Chiavassa et al. 2010). In this context, the strong asymmetry in Zeeman signatures we obtained on that Mira 
would be consistent with a magnetic field tightly linked to the shock wave propagation. 

\section{Discussion}

\subsection{Magnetic fields within evolved stars: observational evidence and theoretical predictions}

The morphology of the CSE of an AGB star will severely change during the quick transition from AGB to PN; the quasi-spherical object becomes axisymmetrical, point symmetrical, or even shows more high-order symmetries (e.g., Sahai \& Trauger 1998).   The classical or generalized interacting stellar winds models (see Kwok, 2000) try to explain this shaping, but 
they have serious difficulties in producing complicated structures with peculiar jets or ansae, and they do not  address the origin of the wind fully. For instance, Bujarrabal et al. (2001) have shown that, in 80 \% of their sample of post-AGB stars, the fast molecular flows have momentum that is too high  (1\,000 times higher in some cases) to be powered by radiation pressure alone, while the magnetic field would help to rule the mass loss geometry and to shape the PN's morphology (Blackman, 2009). 
Direct observational evidence of magnetic fields around PN (Sabin et al., 2007) and around their AGB and post-AGB progenitors have been recently established (see Vlemmings, 2011 for a complete overview).  \\

Models involving a magnetic field playing a role as a catalyst and as a collimating agent have been developed to explain the morphological changes of an evolved star along its transition from the AGB to the PN stage. The simplest models  (see Soker 1998) are based on weak magnetic fields alone, with a strength of about 1 G at the stellar surface. 
Models with a strength of magnetic field greater than  $\sim$10 G in the inner part of the CSE (Soker \& Zaobi, 2002), or also considering other factors contributing to the shaping of the nebulae (rotation, companion, etc.) have been considered (see Balick \& Franck, 2002). From MHD computations, Pascoli (1997) and Pascoli \& Lahoche (2008, 2010)  postulate that ejection of massive winds by an AGB star could be triggered by magnetic activity present in the degenerated core over the typical mass loss process duration (10$^{4}$ years). According to their computations, a toroidal field (of $\sim$ 10$^{6}$ G) produced by a dynamo mechanism in the core could result in a field strength of a few 10 G on the stellar surface. They  also predict that the magnetic field in the low-density envelope of an AGB would mainly be toroidal and would rapidly decrease throughout the CSE with an 1/r  law. 
More recently, Thirumalai \& Heyl (2012) have investigated the occurrence of a magnetic field within Omi Cet, and its impact on the mass loss. They have constructed a hybrid MHD-dust-driven wind model and their best fit is obtained for a surface magnetic field of about 4 G. Our results on $\chi$ Cyg (magnetic field strength at the stellar surface and variation law across the CSE) appear in good agreement with those theoretical predictions. \\

\subsection{A coherent view of the magnetic field at the surface and in the CSE of $\chi$ Cyg}

As the detection level of our Narval data is around the 1/2 gauss level, 
the presence of a strong magnetic field at the surface of $\chi$ Cyg  -when observed 
around its maximum light of 2012 -  certainly has to be excluded. 
 Very recently, Leal-Ferreira et al. (2013) have used polarization of H$_{2}$O maser emissions
 in a sample of AGB stars to present the new panorama for magnetic field-strength dependence along r,  the radial distance to the star (see their Fig.~5). Extrapolated to the photosphere, the r$^{-3}$  law, which corresponds to a dipole field configuration, would lead to a longitudinal magnetic field  strength of several tens of gauss at the stellar surface.  Moreover, this dependence law does not agree with the magnetic field measurements in the outer parts of the CSE, from OH masers (Rudnitski, 2010) and from CN lines for C-rich objects (Herpin et al., 2009), even when considering their associated large error boxes. 
Our Narval observations of $\chi$ Cyg  
rule out   the r$^{-3}$ law with a dipole field configuration (at least for this S-type Mira star). \\

The different locations (in velocity) of the LSD Stokes V and the LSD Stokes Q \& U profiles (as reported in Figs.~1 \& 2, respectively) probably induce different origins for circular and 
linear polarization, not necessarily fully linked to the only shock wave.  
For the LSD Stokes V profile, this reinforces the (Zeeman) magnetic origin, but an effect of the shock wave onto the production  
of the V signature cannot be excluded. Indeed one can expect  an amplification of a weak stellar magnetic field by the important compression inherent to the radiative nature of  a hypersonic shock wave (Fadeyev \& Gillet, 2001).  This phenomenon could be even more important at the photospheric level and in the lower part of the atmosphere, because the shock reaches  its maximal acceleration there. In Mira stars the pulsational phases $\phi$ 
from $\sim$ 0.8 to 1.2 would be particulary relevant to appreciate this effect.\\

However, the shock cannot be the only ingredient at the origin of the magnetism in Mira stars. Indeed, strong hydrogen emission lines are detected in the spectra of 
Miras during 
about 80 \% of their luminosity period, and those features are known to be formed within the radiative wake of the shock wave (Gillet et al., 1983). 
While the propagation of the shock wave can probably be efficient until $\sim$ 10 au or, equivalently, across the SiO maser region and maybe slightly above,  the shock  
will very probably vanish (due to its important radiative losses) before reaching the H$_{2}$O and OH maser regions. The magnetic field measurements  from these external parts of the CSE (see above) can therefore not be 
connected to the shock. While a common origin for the magnetism detected throughout all the parts of the CSE must prevail, the shock could, however, play the role of a compressive amplification on an existing (probably weak) magnetic field,  at the atmospheric level (and maybe also in the SiO maser region).\\

The origin of the surface magnetism in Mira stars still has to be identified, and its impact on the mass loss process needs to be estimated. Soker (2006) argues that magnetic fields detected in evolved stars do not necessarily play a dominant role in shaping PNs, while the presence of a companion would be a  valuable consideration. He  also considers that locally strong magnetic fields detected in AGB environments could be due to the important convection within the stellar atmosphere able to generate a local dynamo  (Dorch, 2004).  
Our results on $\chi$ Cyg support this hypothesis. Indeed, as already suggested for the non--rotating RSG Betelgeuse (see Petit et al., 2013), 
the origin of the surface magnetism in $\chi$ Cyg may rely - at least partly- on a connection to the photospheric or the atmospheric dynamics, through the  generation of a 
magnetic field from a local dynamo powered by the convection (Dorch \& Freytag, 2003).
Both classes of objects, RSG and Mira stars, host a  very complex atmospheric dynamics and a 
few giant convective cells are expected to be present on their surface (Schwarzschild, 1975 ; Chiavassa et al., 2010). 
Moreover, Herpin et al. (2006) show that (SiO) maser emission from Mira stars tends to have higher circular (and linear) polarization compared to RSG stars, indicating a magnetic field that would be stronger in Miras than in RSG, at least in the inner layers of their CSE. Again, this also points  to the likely role of the shock wave (compressive amplification) in this part of the CSE of Miras.

\section{Conclusion}\label{conclusion}

\noindent Using Narval at TBL, we   performed spectropolarimetric observations of the S-type bright Mira star, $\chi$ Cyg, observed  
near its maximum light of 2012. We reported  detection of a faint polarimetric signal in Stokes V spectra, 
and we  established its magnetic origin  (Zeeman effect).  This represents the first detection of a weak magnetic field on the  surface of a Mira star. 
We   estimated   the longitudinal magnetic field on the surface of $\chi$ Cyg to a few gauss, in good agreement with theoretical predictions for the strength of a surface field in an AGB.\\

 Considering the detection level of our data (around the 1/2 gauss level), magnetic field strength of the order of 10 G (or more) at the stellar surface can hardly  be considered. 
Thus, for this S-type Mira star, the hypothesis of a dipole field (inducing a strength of several tens of  gauss on the surface) and a $1/r^{3}$ variation law across the CSE  seems to be excluded (and a  $1/r^{2}$ variation law with a polodial field configuration is also not  favored by our results). On the contrary, our results are in good agreement with the extrapolation 
toward the photosphere of  a $1/r$ variation law of the magnetic field strength across the circumstellar envelope of an AGB. In the future, ALMA, with its full polarimetric abilities, will surely unveil the behavior of the magnetic field in the environment of cool  evolved stars, since it will help remove  geometric effects that may likely affect all the available observations  along the line of sight.\\

The Stokes V signature we  reported on $\chi$ Cyg data also reveals a link with the atmospheric shock wave that is present in the lower part of the atmosphere around the 
phases of the maximum light. We suggest, as already reported for RSGs, that a local dynamo powered by the convection may be a likely explanation for this  surface magnetism. Moreover, the propagation  of the shock wave could probably induce an effect of compressive amplification on a weak field.  To  investigate this  point, a 
monitoring of Mira targets along a part of their pulsation cycles is in progress. Spectropolarimetry at the sub-gauss level also appears mandatory to reveal any weak  
surface fields in Miras. In the future, the near-infrared spectropolarimeter SPIRou will be  perfectly suitable for this kind of study, because it will allow the coverage of the maximum emission 
region of the spectrum of those evolved stars.\\

For the first time in a Mira star observed near its maximum light, we have also reported strong signatures in the LSD Stokes Q and U profiles, associated to metallic lines. 
These features are strong, since they are detected from single sequences (while a combination of 174 V spectra is needed to obtain the circular polarization signal).   
The position of these striking Stokes Q and U profiles, centered on the zero velocity, is  connected  well to the shock front position. These linear polarization signatures reveal that 
 during its propagation in the lower part of the atmosphere, the shock induces  the radial direction on particles as a peculiar geometry. \\

Finally, the presence of a (surface) magnetic field has to be taken into account in the modelization of the atmosphere of a Mira, and also in stellar structure modelization of cool and 
evolved stars. If this magnetic field  plays a dominant role (or a less significant one) in the shaping of the nebula during its evolution after the AGB stage is another point that 
continues to be  debated.

\begin{acknowledgements}
This research has made used of the following services: the SIMBAD database, operated at the CDS (Strasbourg, France), 
NASA's Astrophysics Data System, and the Vienna Atomic Line Database (VALD). We acknowledge with thanks the variable star observations 
from the American Association of Variable Stars Observers (AAVSO) International Database contributed by observers worldwide and used in this research.
We thank the TBL team for providing service mode observations with Narval. The spectropolarimetric data were reduced 
and analyzed using, respectively, the data reduction software Libre-ESpRIT and the Least Squares 
Deconvolution routine (LSD), written and provided by J.-F. Donati from IRAP-Toulouse (Observatoire Midi-Pyr\'en\'ees).
We thank the French Programme National de Physique Stellaire (PNPS) of INSU/CNRS for financial support. 
\end{acknowledgements}

\bibliographystyle{aa}

\end{document}